\documentclass[twocolumn]{pasj00}
\Received{$\langle$reception date$\rangle$}
\Accepted{$\langle$acception date$\rangle$}
\Published{$\langle$publication date$\rangle$}
\SetRunningHead{Michiko S. FUJII}{Formation of young massive clusters}

\usepackage{times,bm}
\usepackage{natbib}
\newcommand {\apgt} {\ {\raise-.5ex\hbox{$\buildrel>\over\sim$}}\ }
\newcommand {\aplt} {\ {\raise-.5ex\hbox{$\buildrel<\over\sim$}}\ } 

\begin{document}

\title{Formation of young massive clusters from turbulent molecular clouds}
\author{Michiko S. FUJII%
\thanks{NAOJ Fellow}}
\affil{%
  Division of Theoretical Astronomy, \\
   National Astronomical Observatory of Japan, 2--21--1 Osawa,\\
   Mitaka-shi, Tokyo 181--8588}
\email{michiko.fujii@nao.ac.jp}
\KeyWords{galaxies: star clusters: general --- open clusters ans associations: general --- open clusters and associations: individual(NGC 3603, Westerlund 1, Westerlund 2, Arches, Trumpler 14) --- methods: numerical}

\maketitle


%%\keywords{open clusters: general --- open clusters: individual(NGC 3603, Westerlund 1, Westerlund 2, R136)}

\begin{abstract}
Young massive clusters are as young as open clusters but 
more massive and compact compared with typical open clusters.
The formation process of young massive clusters is still unclear,
and it is an open question whether the formation process is the
same as typical open clusters or not.
We perform a series of $N$-body simulations starting from initial 
conditions constructed from the results of hydrodynamical simulations
of turbulent molecular clouds. In our simulations, both open clusters
and young massive clusters form when we assume a 
density-dependent star formation efficiency. 
We find that a local star formation efficiency higher than 
50\,\% is necessary for the formation of young massive clusters, but 
open clusters forms from less dense regions with a local star 
formation efficiency of $<50$ \%. 
We confirm that the young massive clusters formed in our simulations 
have mass, size, and density profile similar to those of observed 
young massive clusters such as NGC 3603 and Trumpler 14. We also find
that these simulated clusters evolve via hierarchical mergers of sub-clusters 
within a few Myr, as is suggested by recent simulations and observations. 
Although we do not assume initial mass segregation, 
we observe that the simulated massive clusters show a shallower slope 
of the mass function ($\Gamma\sim-1$) in the cluster center compared
to that of the entire cluster ($\Gamma\sim-1.3$). 
These values are consistent with those of some young massive clusters
in the Milky Way such as Westerlund 1 and Arches.
\end{abstract}

\section{Introduction}

Young massive clusters (YMCs) such as NGC3603, Westerlund 1 and 2
in the Milky Way (MW) and R136 in the Large Magellanic Cloud 
have been found, and their detailed structures have been studied in 
the past ten years. YMCs are as young as open clusters 
($\aplt20$Myr) but as dense as globular clusters
($\sim 10^4M_{\odot}{\rm pc}^{-3}$). The typical mass
of YMCs in the MW is $\sim 10^4M_{\odot}$, which is slightly less massive than
typical globular clusters \citep{2010ARA&A..48..431P}.
They populate mainly in the Galactic disk, and some of them 
(e.g., Arches and Quintuplet) are located close to the Galactic center.
Thus, YMCs seem to be a population different from typical open clusters,
but it is still unclear whether their formation process is the same as
that of open clusters or not.

In our previous papers, we performed a series of $N$-body simulations and 
found that a formation scenario via hierarchical mergers of sub-clusters 
is preferable for YMCs  
in order to explain their dynamically mature characteristics 
(core collapse, mass segregation, and runaway stars)
\citep{2012ApJ...753...85F,2013MNRAS.430.1018F}.
In these studies, however, we adopted simple but artificial 
initial conditions which are consist of 
merging sub-clusters with a mass of $\sim 5000M_{\odot}$. 
In contrast to such a simple model of $N$-body simulations,
observed star forming regions always show more complicated 
clumpy and filamentary structures \citep{2009ApJS..184...18G,
2012A&A...540L..11S,2014Sci...344..183K}.

In order to model more realistic initial conditions for $N$-body simulations,
several methods have been attempted. Hydrodynamical simulations of collapsing 
molecular clouds including  star formation and stellar feedback 
\citep[e.g.,][]{2003MNRAS.343..413B,2008MNRAS.389.1556B,2012MNRAS.419.3115B,
2012ApJ...761..156F,2014ApJ...790..128F,2014MNRAS.442..694D} are the most 
straightforward approach. 
Such simulations, however, require us huge simulation costs to resolve 
individual star formation. Therefore, no simulation large enough for 
the formation of massive clusters with a mass of $\sim 10^4M_{\odot}$
has been performed yet. Compared with such hydrodynamical simulations,
pure $N$-body simulations is numerically much cheaper, and therefore 
it enables us to treat a large number of stars. However, the initial 
distribution of stars we should adopt is unclear.   
\citet{2009ApJ...700L..99A,2010MNRAS.407.1098A} adopted fractal initial 
conditions, which naturally form clumpy structures similar to observed
star forming regions. For the velocity structure of the 
obtained system, they therefore gave 
a random velocity distribution locally following a Gaussian.

An intermediate approach was attempted by \citet{2010MNRAS.404..721M}. They
first performed a hydrodynamical simulation of a turbulent molecular 
cloud which includes star formation using sink particles. After about one 
free-fall time of the initial gas cloud, they stopped the hydrodynamical 
simulations, removed all residual gas assuming instantaneous gas expulsion,
and continued pure $N$-body simulations using a direct $N$-body code. 
This kind of approach seems to work successfully. In 
\citet{2013arXiv1309.1223F}
we adopted a similar method. We also performed a hydrodynamical
simulation using a smoothed-particle hydrodynamics (SPH) method but with 
a lower resolution in order to avoid using sink particles. 
After about one free-fall time, we stopped the hydrodynamical simulation 
and replaced gas particles to stellar particles assuming a star-formation
efficiency depending on the local gas density. Then, we removed all 
residual gas particles and continued a pure $N$-body simulation using a 
direct $N$-body code. With this method, we can construct initial conditions 
for $N$-body simulations based on turbulent molecular clouds and also
more massive than previous works.

In \citet{2013arXiv1309.1223F}, we demonstrated that an ensemble
of star clusters obtained from our simulations successfully reproduce 
observed ones in Carina region and nearby ($<1$kpc from the Sun). 
The most massive clusters formed in the simulations had a mass of 
$\sim 10^4M_{\odot}$, which is as massive as YMCs in the MW.
In this paper, we investigate the detailed formation and evolution 
of YMCs in our simulations and compare the structures 
to those of observed YMCs and open clusters in the MW. 

\section{Simulations}

Our simulations consist of following three steps: 1) hydrodynamical
simulations of turbulent molecular clouds, 2) star formation following
the gas density obtained from the hydrodynamical simulations, and 3)
pure $N$-body simulations of stars based on the structure of the 
molecular clouds. 
 
We perform hydrodynamical simulations using an SPH code,
{\tt Fi} \citep{1989ApJS...70..419H,1997A&A...325..972G,2004A&A...422...55P,
  2005PhDT........17P}, with the Astronomical Multipurpose Software 
Environment (AMUSE)
\citep{2013CoPhC.183..456P,2013A&A...557A..84P}. 
For initial conditions of molecular clouds, we follow those in 
\citet{2003MNRAS.343..413B} and adopt an isothermal
homogeneous gas sphere with a divergence-free random Gaussian
velocity field $\delta \bm{v}$ with a power spectrum 
$|\delta v|^2\propto k^{-4}$ \citep{2001ApJ...546..980O}. 
The spectral index of $-4$ appears in the case of compressive
turbulence (Burgers turbulence), and recent observations of molecular 
clouds \citep{2004ApJ...615L..45H} and numerical simulations 
\citep{2010A&A...512A..81F, 2011ApJ...740..120R,2013MNRAS.436.1245F} 
also suggested values similar to $-4$.
We adopt the total gas mass and size of $4.1\times 10^5M_{\odot}$ and 10 pc,
respectively, which give the mean density of $\sim 1700\,{\rm cm}^{-3}$ 
($\sim 100M_{\odot}{\rm pc}^{-3}$) assuming that the mean weight per particle 
is $2.33m_{\rm H}$. As a consequence, the free-fall time ($t_{\rm ff}$) of 
the gas cloud is 0.83 Myr. 
We set the gas temperature of 30K and zero total energy (potential
plus kinetic energy is zero).
We adopt the mass of SPH particles of $1M_{\odot}$, which is equal to 
the mean stellar mass of the following $N$-body simulations. The 
gravitational softening length is 0.1 pc. We set the number of 
particles within a smoothing length as 64. With these settings, the 
smallest length scale we can resolve is $\sim 0.6$ pc, 
which is smaller than the typical size of embedded clusters (1 pc) 
\citep{2003ARA&A..41...57L} but slightly larger than the observed 
typical width of gas filaments ($\sim 0.1$ pc)
\citep{2011A&A...529L...6A}. In this simulation, we cannot resolve
individual star formation but dense gas clumps in which star 
formation is expected to occur.

After $0.9t_{\rm ff}$, we stop the hydrodynamical simulations and 
analyze the density distribution of the collapsed molecular cloud.
We replace the densest SPH particles with stellar particles by
adopting the local star formation efficiency (SFE), $\epsilon_{\rm loc}$, 
which depends on the local gas density $\rho$ given by 
\begin{eqnarray}
  \epsilon_{\rm loc} = \alpha_{\rm sfe} 
                       \sqrt{\frac{\rho}{100\, (M_{\odot}{\rm pc}^{-3})}},
\label{eq:eff}
\end{eqnarray}
where $\alpha_{\rm sfe}$ is a coefficient which controls the star formation
efficiency and a free parameter in our simulations.
This SFE is motivated by the recent
result that the star formation rate scales with free-fall time
\citep{2012ApJ...745...69K,2013MNRAS.436.3167F}.
We adopt $\alpha_{\rm sfe}=0.02$ in order to reproduce observed SFEs.
In \citet{2012ApJ...745...69K}, they obtained that the SFE per 
$t_{\rm ff}$ is 0.015.
With these settings, we obtain the global SFE for the 
entire regions of several per cent and the SFE of dense regions 
($>1000M_{\odot}{\rm cm}^{-3}$)
of 20--30 \%, which is consistent with observations 
\citep{2003ARA&A..41...57L} and also simulations of 
molecular clouds with star formation \citep{2013ApJ...763...51F}.
The global SFE ($\epsilon$) and the SFE of the dense region ($\epsilon_{\rm d}$)
for the individual models are summarized in Table \ref{tb:models} as model A.
We perform three runs with different random seeds for the turbulence 
for model A, and they are named as A1, A2, and A3.
After replacing gas particles to stellar particles, we randomly assign 
stellar masses following the Salpeter mass function 
\citep{1955ApJ...121..161S} with a lower and upper cut-off mass of 
0.3 and 100$M_{\odot}$. The mass does not conserve locally, but conserve
globally.
The positions and velocities of the stellar particles are identical
to those of the SPH particles. 

We also adopt two models with a constant SFE only for dense regions 
for a comparison. 
One is a constant SFE of 0.8 for the regions with $\rho >5000M_{\odot}{\rm pc}^{-3}$
(model B), and the other is a a constant SFE of 0.3 for the regions with 
$\rho >1000M_{\odot}{\rm pc}^{-3}$ (model C).
While model C follows an assumption that the SFE
cannot exceed 50 \%, model B is motivated by recent numerical study
that in sub-cluster on a length-scale of 0.1--0.2 pc the fraction of 
gas is less than 10\% \citep{2012MNRAS.419..841K,
2012MNRAS.425..450M}.
For both models, the values of 
the total stellar mass, $\epsilon$, and $\epsilon_{\rm d}$
are very similar to those of model A.
These models are also summarized in Table \ref{tb:models}.
We perform two runs with different random seeds for the 
turbulence for model B (B1 and B2) and C (C1 and C2).
A1, B1, and C1 and A2, B2, and C2 have the same random seed, 
respectively. 
In addition, we adopt a model the same as model A but with a
total gas mass of $10^6M_{\odot}$. We call this model as A-1M.

After replacing gas particles to stellar particles, we remove 
all residual gas particles assuming an instantaneous gas expulsion
and perform pure $N$-body simulations with only stars up to 10 Myr
(but 4 Myr for model A-1M because of the calculation cost). For 
$N$-body simulations, we adopt a sixth-order Hermite scheme 
\citep{2008NewA...13..498N} without any gravitational softening. We include 
stellar collisions with sticky sphere approximation with stellar radius 
of zero-age main-sequence following the description of 
\citet{2000MNRAS.315..543H}. We also include stellar mass-loss at 
the end of the main-sequence life time following \citet{2000MNRAS.315..543H}
\citep[see][for the details]{2009ApJ...695.1421F, 2013MNRAS.430.1018F}.  
We use a time step criterion of \citet{2008NewA...13..498N} with an accuracy
parameter, $\eta=$0.1--0.25.  The energy error was $\aplt 10^{-3}$ for
all simulations over the entire duration of the simulations.

\begin{table*}
\begin{center}
\caption{Initial conditions\label{tb:models}}
\begin{tabular}{ccccccccccc}\hline \hline
Model & $M_{\rm gas}$ & $r_{\rm gas}$ & $\rho_{\rm gas} $ & SFE & $\rho_{\rm c} $ & $M_{\rm star}$ & $N_{\rm star}$ & $\epsilon$ & $\epsilon_{\rm d}$ & $Q$ \\ 
  & $(M_{\odot})$ &  (pc) & ($M_{\odot}{\rm pc}^{-3}$) &  & ($M_{\odot}{\rm pc}^{-3}$) &  $(M_{\odot})$ &  &  & & \\\hline
A1 & $4.1\times 10^5$ & 10 & 100 & eq.(\ref{eq:eff}) & - &  $3.2\times 10^4$  & 31895 & 0.078 & 0.22 &  2.6\\ %s1
A2 & $4.1\times 10^5$ & 10 & 100 & eq.(\ref{eq:eff}) & - & $4.3\times 10^4$ & 42596 & 0.096 & 0.25 & 0.85 \\ %s3
A3 & $4.1\times 10^5$ & 10 & 100 & eq.(\ref{eq:eff}) & - & $2.3\times 10^4$ & 23273 & 0.057 & 0.16 & 8.4 \\ %s4
B1  & $4.1\times 10^5$ & 10 & 100 & $0.8$& $5\times 10^3$ & $3.4\times 10^4$ & 33974 & 0.083 & 0.30 & 1.6 \\
B2  & $4.1\times 10^5$ & 10 & 100 & $0.8$& $5\times 10^3$ & $4.3\times 10^4$ & 42710 & 0.10 & 0.30 &  0.84\\
C1  & $4.1\times 10^5$ & 10 & 100 & $0.3$& $10^3$ &  $3.4\times 10^4$ & 34086 & 0.084 & 0.30 & 4.2\\
C2  & $4.1\times 10^5$ & 10 & 100 & $0.3$& $10^3$ &  $4.4\times 10^4$ & 43500 & 0.11 & 0.30 & 2.9\\
A-1M & $1\times 10^6$ & 13.4 & 100 & eq.(\ref{eq:eff}) &  -     & $1.1\times 10^5$ & 109080 & 0.11 &  0.27 & 1.8\\
\hline
\end{tabular}
\medskip
\\$M_{\rm gas}$, $r_{\rm gas}$, $\rho_{\rm gas}$ are the mass, radius, and density of the molecular cloud,
respectively.
SFE and $\rho_{\rm c}$ indicate the SFE and the critical density for the constant SFE. 
$M_{\rm star}$ and $N_{\rm star}$ are the total mass and number of stars for $N$-body simulations.
$\epsilon$ and $\epsilon_{\rm d}$ are the SFE for the entire system and the region with a local
density of $>1000M_{\odot}{\rm pc}^{-3}$, respectively.
$Q$ is the virial ratio of stellar particles at the beginning of $N$-body simulations. If $Q=1.0$, the system is in virial equilibrium. 
\\
\end{center}
\end{table*}

\begin{figure*}
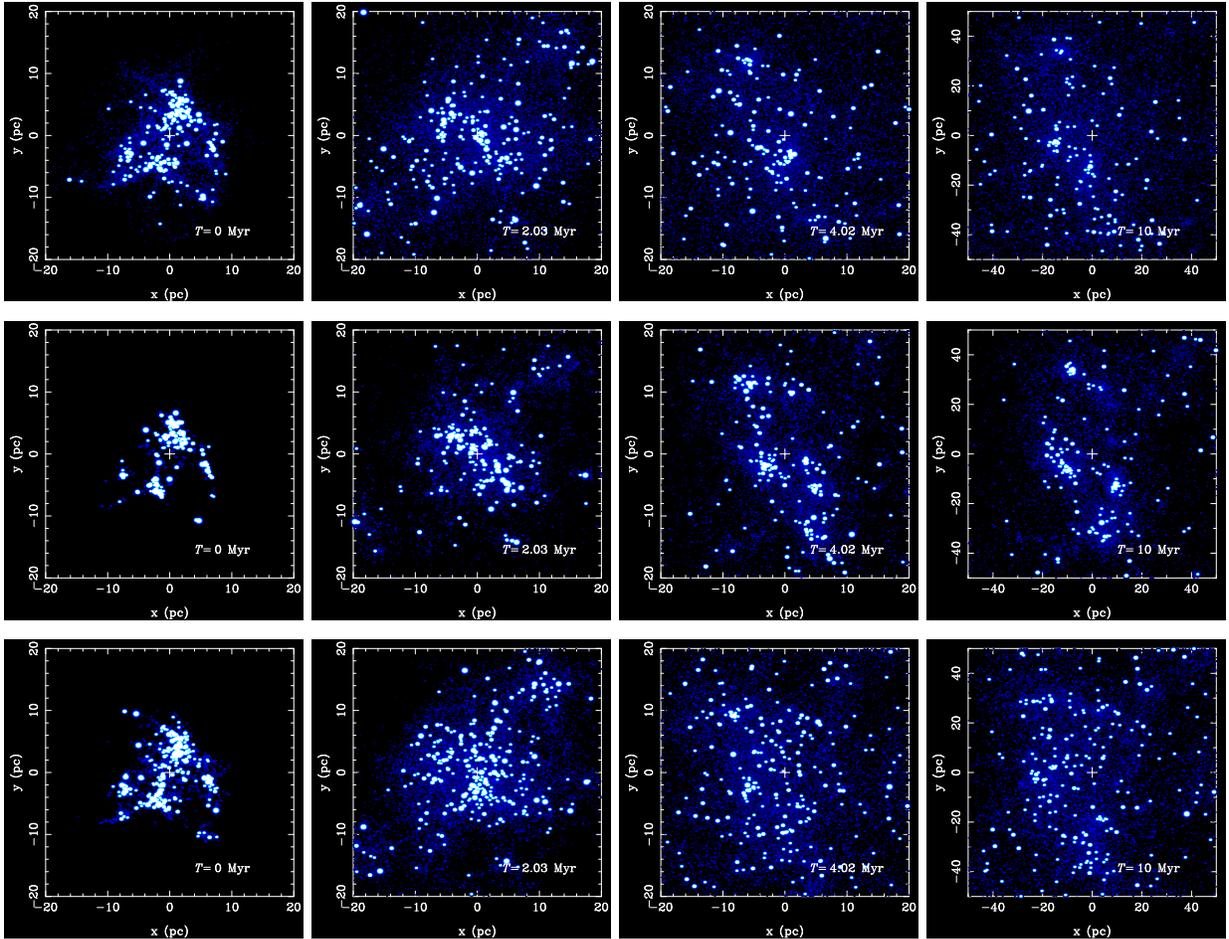

\begin{center}
\FigureFile(40mm,40mm){f1a.eps}
\FigureFile(40mm,40mm){f1b.eps}
\FigureFile(40mm,40mm){f1c.eps}
\FigureFile(40mm,40mm){f1d.eps}\\
\FigureFile(40mm,40mm){f1e.eps}
\FigureFile(40mm,40mm){f1f.eps}
\FigureFile(40mm,40mm){f1g.eps}
\FigureFile(40mm,40mm){f1h.eps}\\
\FigureFile(40mm,40mm){f1i.eps}
\FigureFile(40mm,40mm){f1j.eps}
\FigureFile(40mm,40mm){f1k.eps}
\FigureFile(40mm,40mm){f1l.eps}\\
\end{center}
\caption{Snapshots at $t=$ 0, 2, 4, and 10 Myr for models A1, B1, and C1
from top to bottom. \label{fig:snap}}
\end{figure*}

\section{Formation and Evolution of Star Clusters}

In Figure \ref{fig:snap} we present the snapshots
at $t=$ 0, 2, 4, and 10 Myr for models A1, B1, and C1.
Hereafter, we set the beginning of $N$-body simulations as 
$t=0$.  
Since these models have the same random seed for the turbulence, the 
initial distributions of stars are very similar. 
At $t=2$Myr, some massive star clusters form especially 
in model A1 and B1.  In model C1, the final distribution of stars 
is more elongated and the formed clusters are smaller than
those in model A1 and B1.
In this section, we detect the clusters 
formed in our simulation and investigate their time evolution.

\subsection{Mass-Radius Diagram of Star Clusters}
At $t=0.5$, 2, 4, and 10 Myr, we interrupt the simulations and
detect star clusters using a clump finding method, HOP 
\citep{1998ApJ...498..137E}, in the AMUSE framework. 
We adopt a parameter as following:
the outer cut-off density of $\rho_{\rm out} =
4.5M_{\rm s}/(4\pi r_{\rm h}^{3})$ (three times the half-mass
density of the entire system, $\rho_{\rm h}$), the saddle density threshold of $8\rho_{\rm out}$,
the peak density threshold of $10\rho_{\rm out}$, the number of particles for 
neighbor search of $N_{\rm dense}=64$, the number 
of particles to calculate the local density of $N_{\rm hop}=64$, 
and the number of particles of neighbors to determine for two groups to merge
of $N_{\rm merge}=4$. With this set-up, the detection limit of the
clump mass is $\sim 100M_{\odot}$.
Since HOP sometimes detected multiple clumps as one clump, we apply the HOP
method again for the detected clumps if the half-mass density of
the clump ($\rho_{\rm h, c}$) is less than $100\rho_{\rm h}$ adopting
$\rho_{\rm out} = \rho_{\rm h, c}$. If $\rho_{\rm h, c}>100\rho_{\rm h}$, 
such a clump is compact and therefore does not seem to contain
multiple clumps.
We confirmed that all detected clumps do not contain multiple clumps. 

\begin{figure*}
\begin{center}
\FigureFile(70mm,70mm){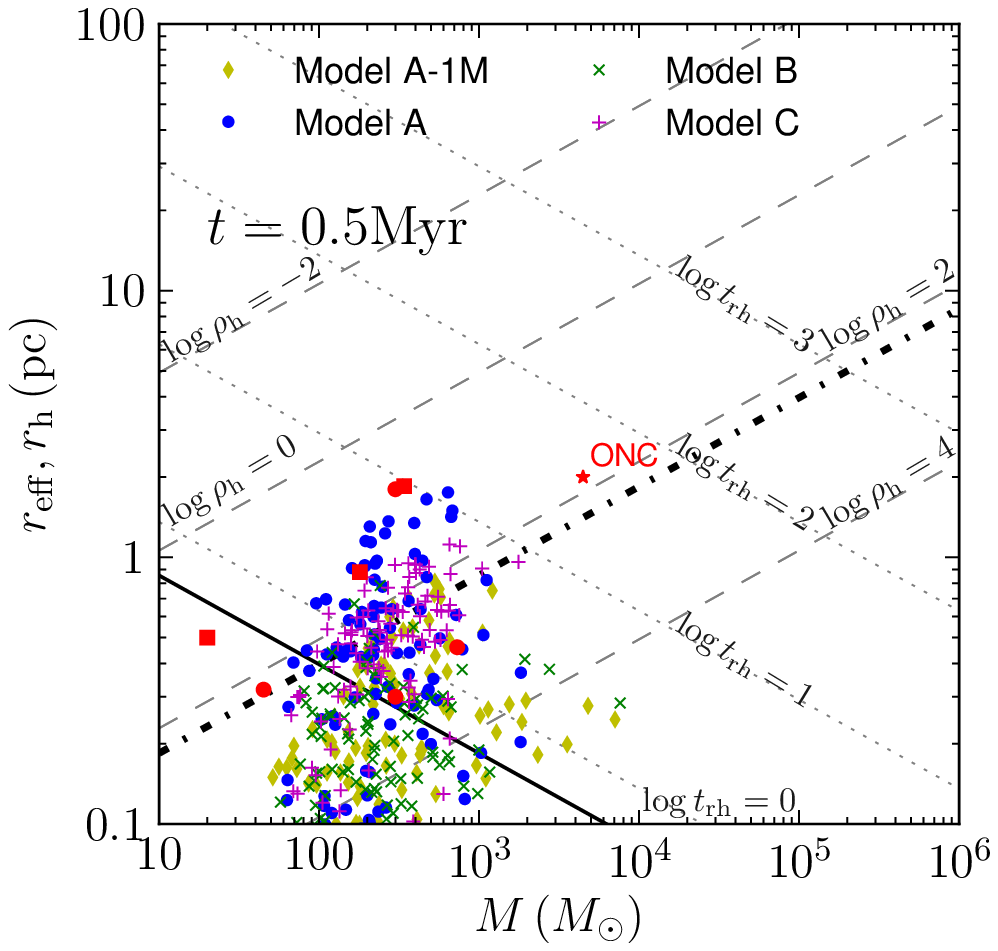}
\FigureFile(70mm,70mm){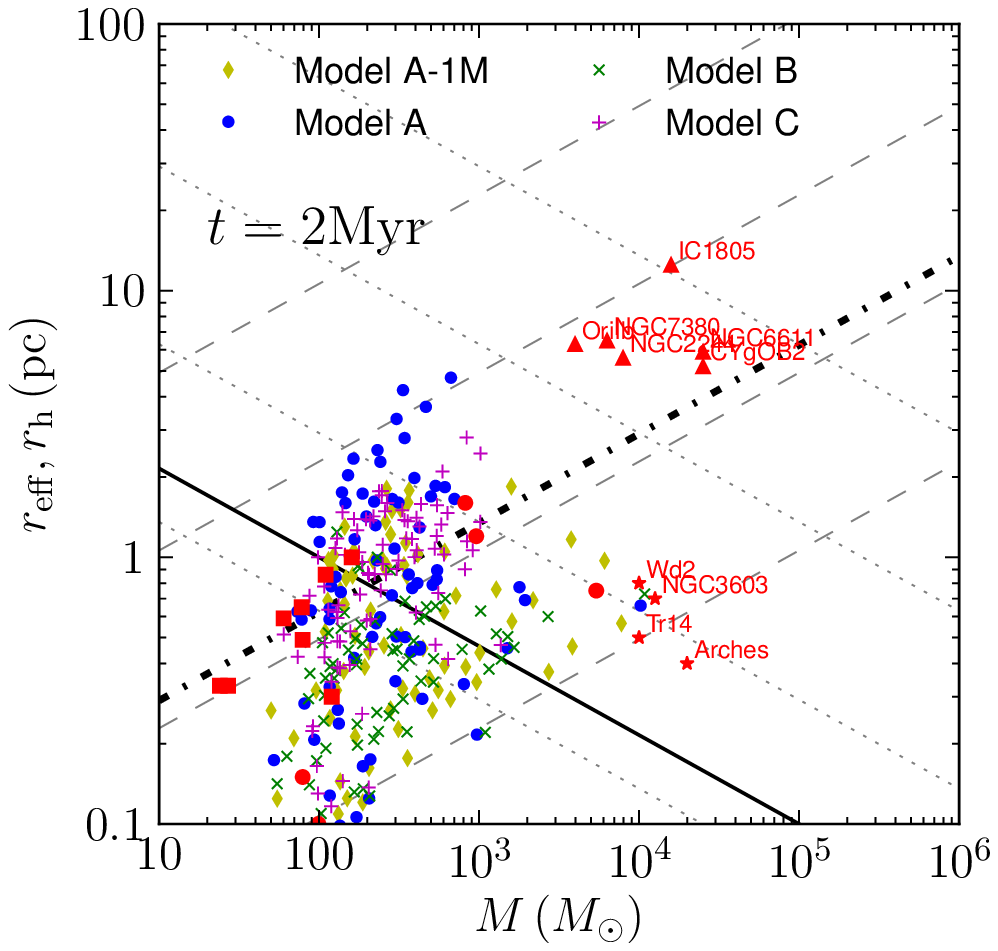}
\FigureFile(70mm,70mm){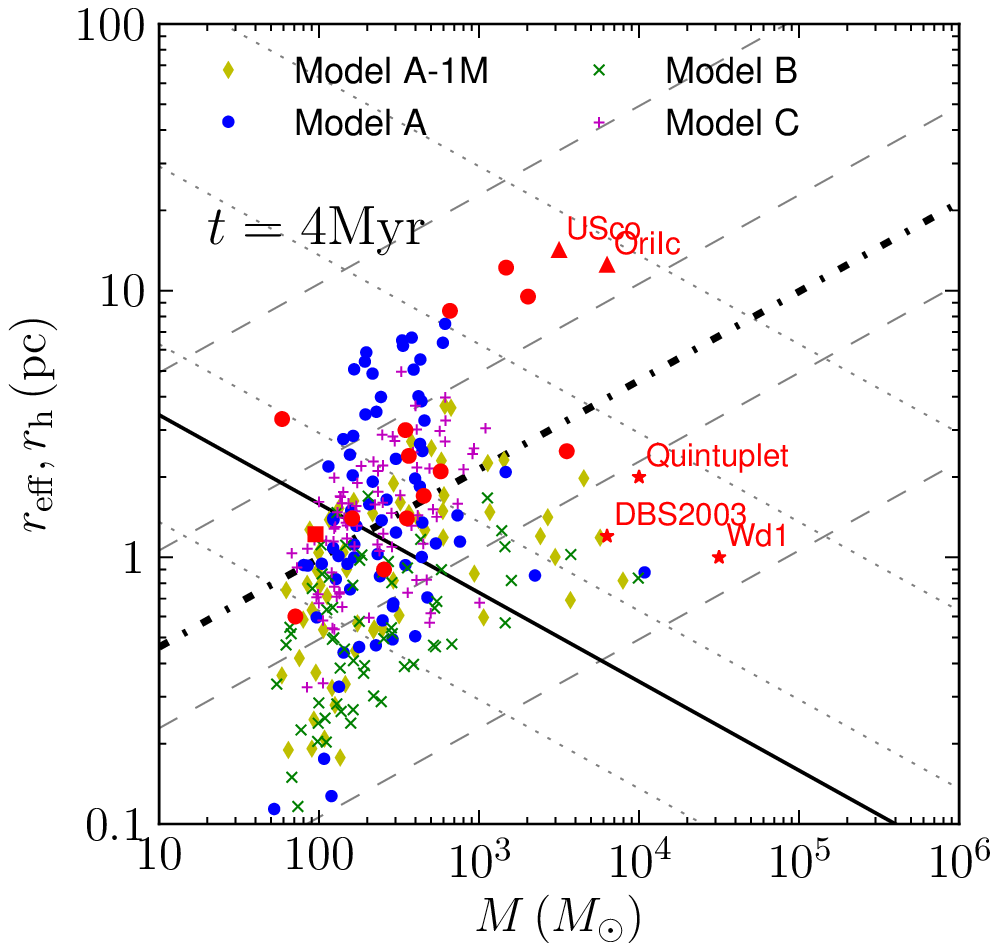}
\FigureFile(70mm,70mm){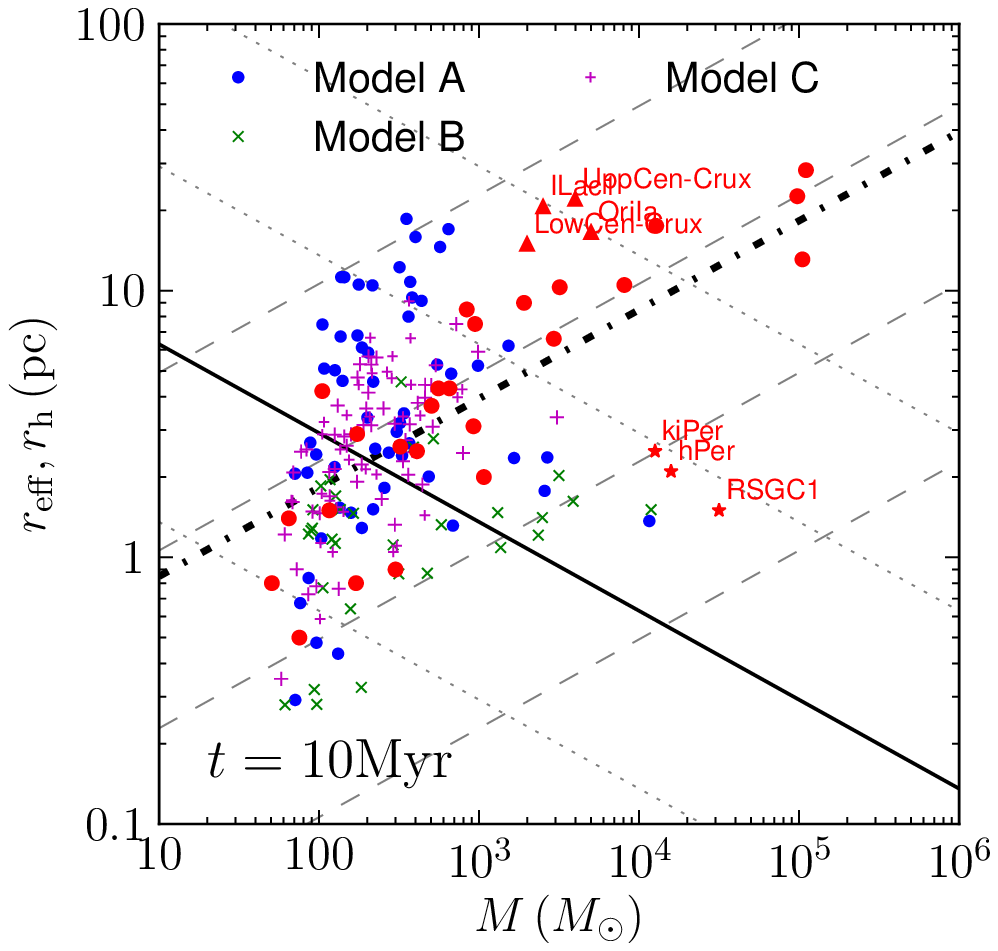}
\end{center}
\caption{Mass-radius diagram of clusters obtained from the simulations 
  at 0.5, 2, 4, and 10 Myr. Same models with different random seeds are 
  plotted with the same symbols.
  Red points indicate observed MW clusters with ages of 
  $\leq 1$Myr, 1--3 Myr, 3--6 Myr, 6--15 Myr.
  Squares, circles, stars, and triangles are for embedded, 
  open, young massive \citep{2010ARA&A..48..431P}
  and leaky clusters, respectively.
  The data is from \citet{2003ARA&A..41...57L, 2008A&A...487..557P, 2009A&A...498L..37P,
  2010ARA&A..48..431P,2009AJ....137.4777W,2003ApJ...593.1093L,2006AJ....132.2296A,
  2009A&A...504..461F,2006ApJ...646.1215L,2008AJ....135..966F,2011MNRAS.414.3769B,
  1997AJ....113.1788H,1997MNRAS.286..538D,1991AJ....102.1108H}. We assumed that 
  the age of embedded clusters given in \citet{2003ARA&A..41...57L} is 1 Myr. 
  The observed clusters and associations with the names are given in \citet{2010ARA&A..48..431P}.
  Black thick solid and dash-dotted lines indicate the line at which the relaxation 
  time and the dynamical time are equal to the age. 
  \label{fig:mass_radius_age}}
\end{figure*}

In Figure \ref{fig:mass_radius_age} we present the mass-radius diagram 
of the detected clusters at each time. We also plot observed young 
star clusters in the MW. 
Most of simulated clusters are located in the bottom-left region
in this diagram. Clusters in this region are observationally categorized 
as open or embedded clusters. Embedded clusters are very young and, 
therefore, they are still covered with surrounding molecular clouds. 
Typical mass and size are a hundred solar mass and $\sim 1$ pc, 
respectively \citep{2003ARA&A..41...57L}.
The distribution of open and embedded clusters
shifts to the upper region in the diagram with time. 
This is due to the dynamical evolution of star clusters.
\citet{2011MNRAS.413.2509G} modeled the dynamical evolution 
of star clusters based on H{\'e}non's model 
\citep{1961AnAp...24..369H,1965AnAp...28...62H}. After the first
mass segregation (or core collapse) phase, star clusters expand
due to the energy generation in the core. In this expansion phase,
the half-mass relaxation time ($t_{\rm rh}$) increases with the cluster
age ($t_{\rm age}$), and $t_{\rm rh}\sim t_{\rm age}$. 
In the figure, we indicate the half-mass relaxation time 
for a given mass and half-mass radius with gray dotted lines. 
We adopt 
\begin{eqnarray}
t_{\rm rh}\sim 2\times 10^8 \left( \frac{M}{10^6M_{\odot}} \right)^{1/2} \left( \frac{r_{\rm h}}{1{\rm pc}}\right)^{3/2} {\rm year} 
\label{eq:t_rh}
\end{eqnarray}
assuming $r_{\rm vir}\sim r_{\rm h}$ and $\langle m \rangle = 1M_{\odot}$
\citep{2010ARA&A..48..431P}.  
The black thick lines in Figure \ref{fig:mass_radius_age} indicate
the line at which $t_{\rm rh}=t_{\rm age}$. 
Open and embedded clusters are mainly located around this line 
and evolve with it. Such dynamical evolution of open clusters
is also seen in simulations performed by \citet{2012MNRAS.425..450M}.
Since the initial relaxation time of the open and embedded clusters are 
around 1 Myr, they dynamically evolve immediately and shift to the
expansion phase. From the evolution in this diagram, embedded clusters
seems to be ancestors of typical open clusters.

YMCs distribute an area clearly different from that of
open and embedded clusters. They are located at the middle right of 
the mass-radius diagram ($\sim 10^4M_{\odot}{\rm pc}^{-3}$ and $\sim 1$ pc) 
at $t=2$, 4, and 10 Myr. 
In our simulations, we also find a branch of cluster distribution
similar to YMCs, although the number of such dense massive clusters 
are much less compared to the main
distribution of clusters similar to open clusters.
We treat these massive ($\sim 10^4M_{\odot}$) clusters as YMCs and
compare them to observed YMCs in Section 4.

The third population of clusters distribute at the top-right
region in the mass-radius diagram. They are so-called 
``leaky'' clusters \citep{2009A&A...498L..37P} but also 
categorized as association because they seem to be unbound systems 
\citep{2010ARA&A..48..431P}. They are as massive as YMCs but much
less dense. 
In this diagram, especially at $t=10$ Myr, the leaky clusters seem 
to populate at the high-mass end of open and embedded clusters.

\citet{2010ARA&A..48..431P} proposed to use the ratio of the age 
and dynamical time ($t_{\rm dyn}$) of clusters as an indicator of bound 
systems. The dynamical time is given as
\begin{eqnarray}
t_{\rm dyn}\sim2\times 10^4\left( \frac{M}{10^6M_{\odot}} \right)^{-1/2} \left(\frac{r_{\rm h}}{1{\rm pc}}\right)^{3/2}
{\rm year}
\label{eq:t_dyn}
\end{eqnarray}
\citep{2010ARA&A..48..431P} assuming $r_{\rm vir}\sim r_{\rm h}$.
If $t_{\rm dyn}$ is much shorter than the cluster age ($t_{\rm age}$), the cluster 
seems to be bound. If $t_{\rm dyn}$ is comparable to $t_{\rm age}$ or 
even longer, the cluster seems to be unbound.
\citet{2010ARA&A..48..431P} suggested that if $t_{\rm age}/t_{\rm dyn}<3$, 
the system is unbound. With this criterion, they categorized 
the leaky clusters as associations (unbound systems).
The black dash-dotted lines in Figure \ref{fig:mass_radius_age} 
indicate the density with 
which the dynamical time of the cluster is equal to the age. 
Most of the MW open clusters are located near the boundary,
and therefore they seems to be loosely bound or some of them may
not be bound. While most open clusters and embedded clusters 
distribute also around the line with $t_{\rm age}=t_{\rm rh}$, 
leaky clusters and some massive open clusters (they might be a
population the same as the leaky clusters) distribute 
far away from this line. This means that the leaky and typical
open clusters are dynamically in different evolutionary phases.

In our simulation, no cluster similar to the leaky clusters forms. 
Is it possible to form leaky clusters (or associations) in
the same process we adopted? 
The maximum mass of star clusters formed in a turbulent molecular 
cloud is expected to increase with the mass of the molecular cloud
\citep{2013arXiv1309.1223F}.
If we increase the initial gas mass, the distribution of open 
clusters might elongate toward the leaky clusters.
In model A1-1M, which has $10^6M_{\odot}$ of the initial gas mass, 
we found a
larger number of YMCs compared with model A, but the mass of 
open clusters was not much different from those of model A.
Another possibility is that we fail to detect leaky clusters
because of our clump finding method.
\citet{2001ApJ...563..151M} categorized young 
massive stellar systems ($<20$Myr and $>3\times 10^4M_{\odot}$) 
in local galaxies evaluating their morphology using one-quarter, one-half,
and three-quarters light-radii. Some clusters in their sample were 
categorized as ``scaled OB associations,'' which do not have clear 
center but show asymmetric, elongated, and clumpy structures. 
The typical size of the scaled OB associations a few tens pc, 
which is comparable to those of the leaky clusters.  
Some of leaky clusters are also known as sub-clustered system
\citep{2014MNRAS.438..639W}.
Using our clump finding method, such clumpy clusters 
could be detected as some smaller clumps instead of a large
clumpy system.

\subsection{Formation of Young Massive Clusters via Hierarchical Mergers}

\begin{figure}
\FigureFile(70mm,70mm){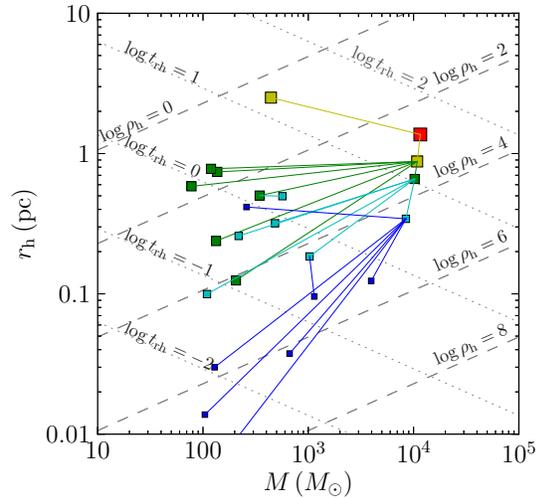}
\caption{Merger history for in mass-radius diagram for the most massive cluster in model A. Blue, cyan, green, yellow, and red squares indicate clusters at 0.25, 0.5, 2, 4 and 10 Myr, respectively. The lines indicate the assembling and the evolution of the clusters. \label{fig:merger_history}}
\end{figure}

As we see in the previous subsection, 
the majority of clusters evolve upward in the 
mass-radius diagram due to the relaxation. Some clusters,
however, especially massive clusters, increases their mass and move 
from left to right in the diagram. 
We find that these massive clusters experienced hierarchical mergers. 
In Figure \ref{fig:merger_history}, we present the merger history
of the most massive cluster in model A. The cluster finally evolved
to a cluster which has a radius and mass similar to those of YMCs 
($\sim 1$ pc and $\sim10^4M_{\odot}$). During 10 Myr, this cluster 
accumulated 17 smaller clusters, but most of mergers occurred before 
4 Myr.

Such a formation scenario via hierarchical mergers is preferable 
for YMCs. 
Observed YMCs show dynamically mature features such as mass segregation 
and the existence of runaway stars, which occur on the relaxation time scale.
The relaxation time obtained from their current mass and size, however,
implies that they are too young to show such dynamically mature features. 
Merger scenario can overcome this discrepancy. Since smaller clusters 
in general have a shorter relaxation time, they can dynamically 
evolve within a few Myr. If such smaller sub-clusters assemble in 
a few Myr and form a massive cluster, the merged cluster shows 
dynamically
evolved features which are taken over from the sub-clusters.
In \citet{2012ApJ...753...85F}
we demonstrated that the degree of mass segregation, the distribution of 
massive stars, and the formation of runaway stars in observed YMCs are
successfully explained by the assembling of sub-clusters. Observationally,
a remnant of sub-clusters was found in the vicinity of R136
\citep{2012ApJ...754L..37S}.

In Figure \ref{fig:merger_history} the first cluster is very dense but
has a short relaxation time less than 1 Myr. If the cluster failed to 
experience merger with other dense sub-clusters, it would evolve 
to a less dense cluster on its relaxation timescale. If the cluster,
however, merged with other dense sub-clusters before
it becomes less dense, the cluster can evolve into a more
massive cluster maintaining its high density. After repeating mergers, 
the relaxation time exceeds 10 Myr at $t=2$ Myr, and therefore 
the dynamical evolution slows down. In section 4, we discuss the 
detailed structure of this massive cluster and compare it with 
observed YMCs.

\subsection{The Effect of the Star Formation Efficiency}

In Figure \ref{fig:mass_radius_age} we see that the distributions 
of detected clusters in our simulation depends on the model for
the SFE; only open clusters form in model C, while YMCs form 
in model B. 
In model A (density-dependent SFE), both populations form 
at the same time. 
In the middle and bottom panels in Figure \ref{fig:snap}, we present 
the snapshots of model B and C. While massive clusters are seen in
model A and B, model C shows a less clumpy structure. 
We perform clump findings at each time and the mass and size of the
obtained clusters are plotted in Figure \ref{fig:mass_radius_age}.
As seen in Figure \ref{fig:snap}, model C does not form any
YMC-type clusters, but in model B the formed clusters mainly distribute 
the YMC brunch.

These differences of cluster formation modes are caused by the local 
SFE we assumed. In model C, all clusters are dominated by gas when 
they form because we adopt a constant SFE of 30\,\%.
These clusters always become less dense after gas expulsion.
In model A or B, however, the densest regions are allowed to be dominated
by stars, and as a consequence, dense clumps can maintain
their high density over gas expulsion. The dense clumps can merge 
with each other before they evolve to less dense clusters 
due to the two-body
relaxation and finally grow up to dense massive clusters. 
With a density-dependent SFE we assumed, the local SFE 
exceeds 50\% when $\rho>6.3\times10^4M_{\odot}{\rm pc}^{-3}$.
This value is roughly consistent with the density
of observed YMCs ($\rho\sim\times10^4M_{\odot}{\rm pc}^{-3}$, 
see Figure \ref{fig:mass_radius_age}). Thus,
we conclude that a local SFE higher than 50\,\% is necessary for the 
formation of YMCs and that a density-dependent SFE (i.e., a constant 
SFE per free-fall time) is preferable to explain the formation of both 
open clusters and YMCs in the same mechanism. 

\section{Properties of YMCs in simulations}

\begin{figure*}
\begin{center}
\FigureFile(70mm,70mm){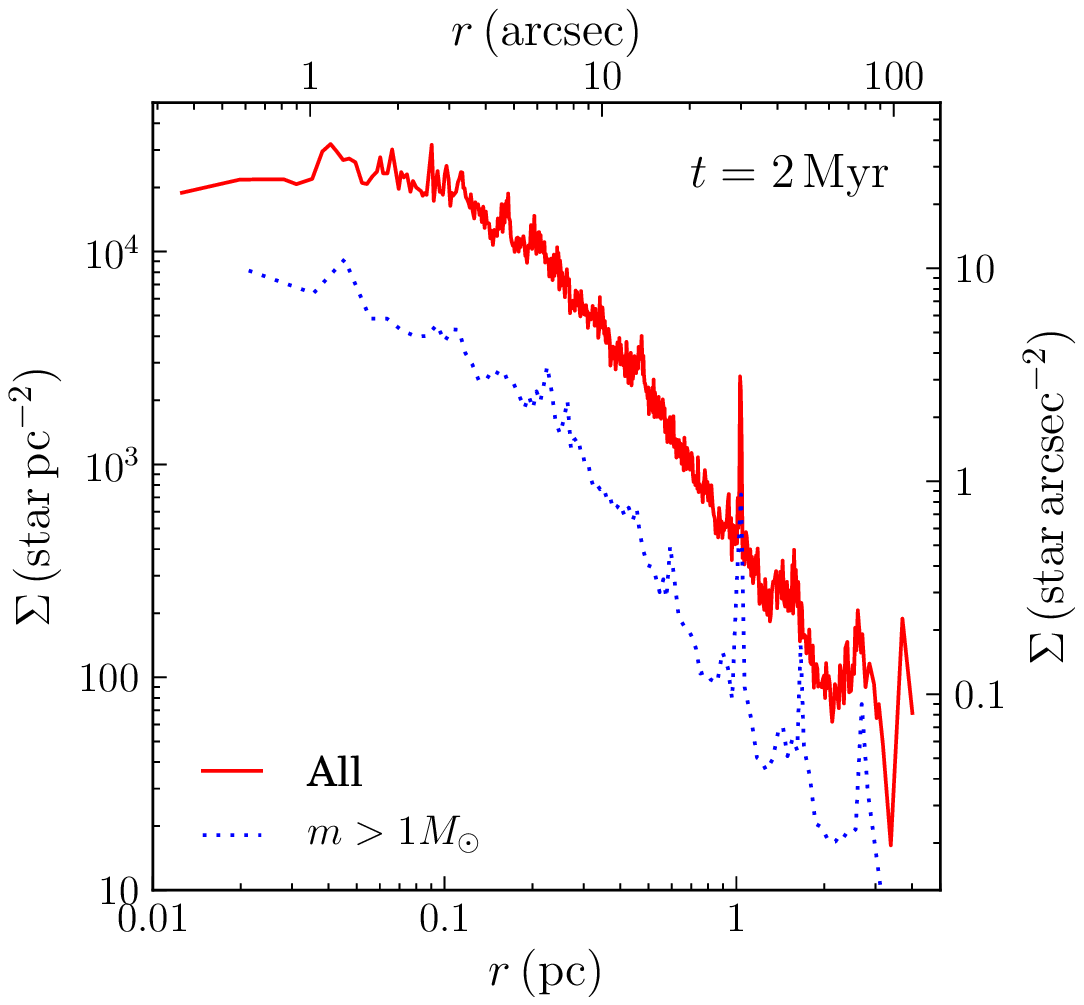}
\FigureFile(70mm,70mm){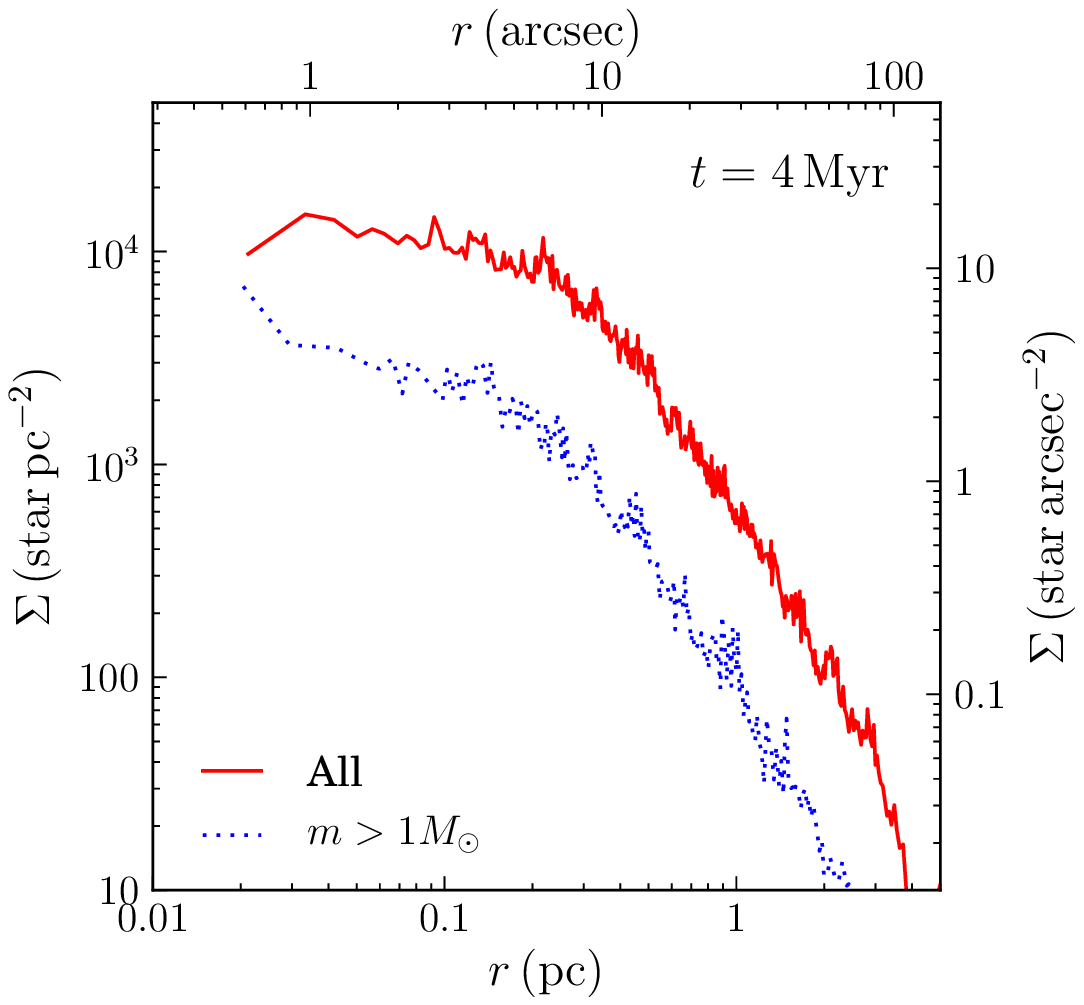}
\end{center}
\caption{Projected stellar density of the most massive cluster in model A
at 2 (left) and 4 (right) Myr.
Right and top axes are in arcsec and star\,arcsec$^{-2}$ assuming that the 
cluster exist the distance of NGC 3603. We assume that $1''=0.035$ pc
\citep{1998ApJ...498..278E}. Red full and blue dashed curves indicate 
the projected density profile for all stars and stars with $>1M_{\odot}$,
respectively. The core radii are 0.21 and 0.19 pc for all and $>1M_{\odot}$
stars, respectively.
\label{fig:sd}}
\end{figure*}

In the previous section, we showed that YMC-like clusters form in
our simulations. Are they really comparable to YMCs in the MW?
In Figure \ref{fig:sd} we present the surface density profiles of 
the largest cluster in model A at 2 and 4 Myr. 
The total mass and the half-mass radius are 
$1.1\times 10^4M_{\odot}$ and 0.88 pc at 2 Myr, respectively, which are 
very similar to those of NGC 3603 ($1.3\times10^4M_{\odot}$ and 0.70 pc)
\citep{2010ARA&A..48..431P}. The spikes at around 1 pc at 2\,Myr are 
sub-clumps merging to the main cluster, and they disappear at 4\,Myr.
We measure the core radius and density using 
a local density method of \citet{1985ApJ...298...80C}.
The original method is applied for 
volume densities, but we here apply it for surface densities.

Using this method, we obtained the core radius of 0.21 pc for all 
stars. We also measured the core radius only for stars with $m>1M_{\odot}$
because less massive stars are not detected in observations and 
obtained a core radius of 0.19 pc. These values are comparable to 
that of NGC 3603: 0.45 pc in \citet{1998ApJ...498..278E}
and 0.15 pc in \citet{2010ARA&A..48..431P}. 
The core densities of the simulated clusters are $1.9\times 10^4$ 
and $4.5\times 10^3$
star\,pc$^{-2}$ for all and $>1M_{\odot}$ stars, respectively. The 
core density for stars with  $>1M_{\odot}$ is also comparable to 
the observed one \citep[see Fig. 8 in][]{1998ApJ...498..278E}.
Trumpler 14 is also a young massive cluster which has a mass, size, 
and age similar to NGC 3603 (see top-right panel of Figure 
\ref{fig:mass_radius_age}). 
The surface density profile is observationally obtained, and the 
radius and surface density of the core are estimated to 
be 0.11--0.22 pc and $4\times 10^3$--$1.1\times 10^4$ star\,pc$^{-2}$,
respectively \citep{2010A&A...515A..26S}. These values are also 
consistent with those obtained from our simulations.
At $t=4$ Myr, the core becomes larger and the density decreases. 
The core density and radius are $9.3\times10^3$ star\,pc$^{-2}$
and 0.31 pc for all stars and $2.6\times10^3$ star\,pc$^{-2}$
and 0.25 pc for stars with $m>1M_{\odot}$. 

\begin{figure*}
\begin{center}
\FigureFile(80mm,80mm){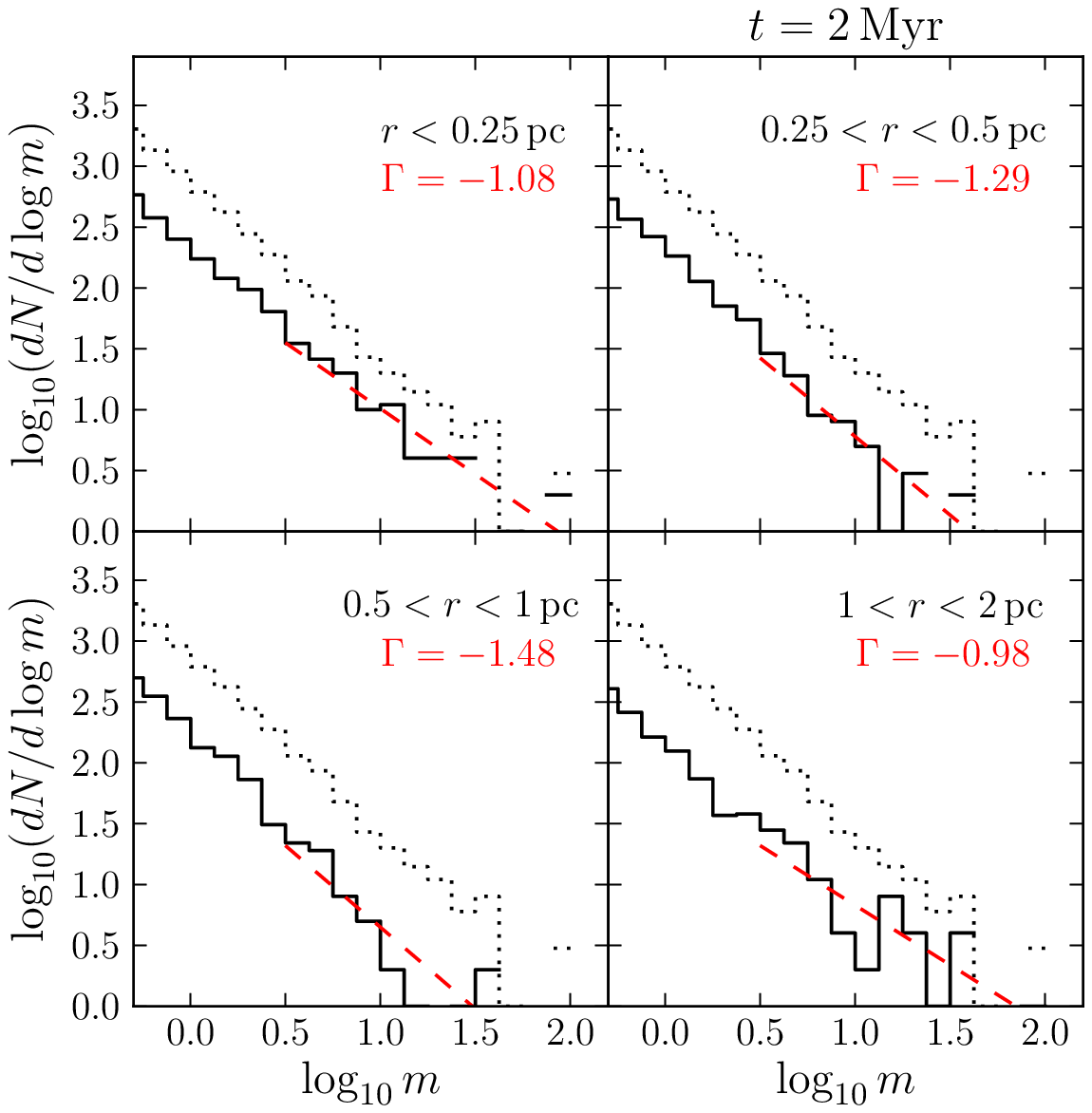}
\FigureFile(80mm,80mm){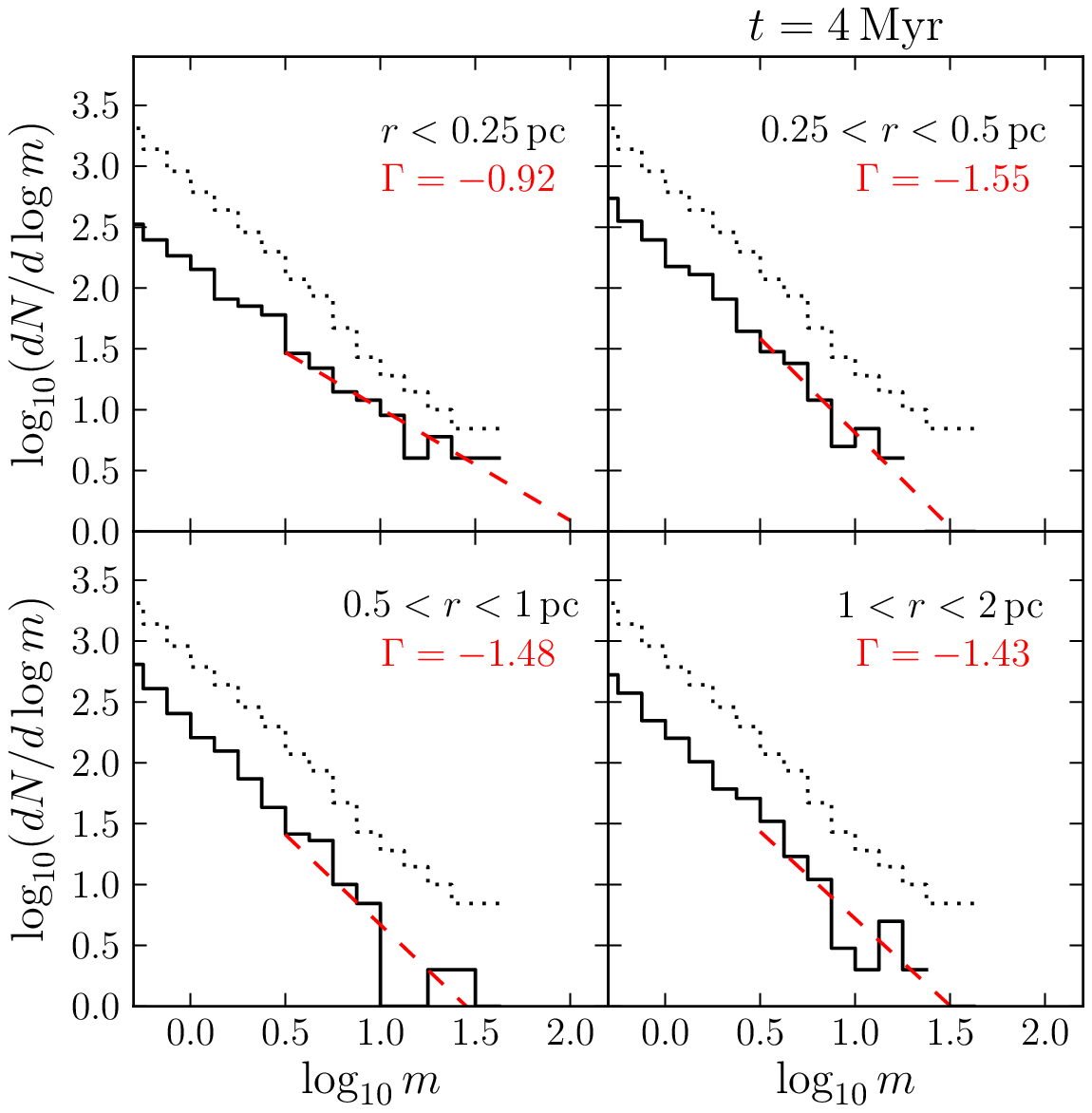}
\end{center}
\caption{Mass function of the most massive cluster in model A in a sequence 
of increasing annuli at 2 (left) and 4 Myr (right). The dotted histogram
is the total mass function within 2 pc. Red dashed line indicates the 
result of a least-square fit for the data points with $\log_{10}m>0.5$.
The slope of the mass function is indicated as $\Gamma$. 
The slope of the total mass function obtained from the fitting is 
$\Gamma=-1.34$ and $-1.31$ for 2 and 4\,Myr, respectively.\label{fig:mf}}
\end{figure*}

Although we did not take into account any initial mass segregation,
the formed clusters are mass segregated as are YMCs. 
In Figure \ref{fig:mf} we present the mass function of the massive 
cluster shown in Figure \ref{fig:sd} in a sequence of increasing annuli
at 2 and 4 Myr. We obtain the slope of the mass 
function ($\Gamma$) using a least-squares fit. For the fitting, we include 
only the data points with $\log_{10}m>0.5$, where $m$ is the stellar mass, 
because most of the observational data are complete down to a 
few solar masses
\citep{2013ApJ...764...73P}. We find that the slope of the mass
function tends to be steeper as the annuli are increasing.
(Note that the power-law slope at $1<r<2$ pc at 2Myr is also shallow, 
but this is caused by a sub-cluster in this annulus.)
This trend is clear especially in the inner most annulus ($r<0.25$pc),
and the slope is $\Gamma\sim-1$ ($\Gamma=-1.35$ for Salpeter mass function,
which is the initial mass function we assume). 
Such shallower mass functions in inner annuli are commonly
observed in YMCs in the MW such as NGC3603 \citep{2013ApJ...764...73P},
Westerlund 1 \citep{2013AJ....145...46L}, and 
Arches \citep{2013A&A...556A..26H}. The values of slopes are different
in each YMC; 
$\Gamma=-0.88$ for total and $\Gamma==-0.26$ for $r\aplt0.2$ pc for NGC3603
\citep{2013ApJ...764...73P},
$\Gamma=-1.4$ \citep{2011MNRAS.412.2469G}
or $\Gamma=-0.8$ in total and $\Gamma=-1.1$ for $r \aplt0.3$ pc
for Westerlund 1 \citep{2013AJ....145...46L},
and $\Gamma=-1.53$ for total and $\Gamma=-0.5$ for $r<0.2$ pc for Arches
\citep{2013A&A...556A..26H}, respectively.

The mass segregation proceeds in a relaxation time. The relaxation time 
obtained from the current masses and sizes of YMCs in the MW are $\sim 10$ Myr 
(see Figure \ref{fig:mass_radius_age}),
which is longer than their age. If these YMCs formed via mergers as is
seen in our simulations, their ancestor sub-clusters would have had a much 
shorter relaxation time of less than 1\,Myr (see Figure 
\ref{fig:merger_history}). 
Since the dynamically mature features remain after mergers 
\citep{2012ApJ...753...85F,2013MNRAS.430.1018F}, the mass segregation 
in the YMCs seems to be a natural consequence of the dynamical evolution
via mergers. 

\section{Summary}
We performed a series of direct $N$-body simulations of young star 
clusters
starting from initial conditions constructed from hydrodynamical
simulations of turbulent molecular clouds. In our simulations,
we obtained an ensembles of young star clusters. Most of them
had a mass and size similar to those of typical open clusters
in the MW. 
In the mass-radius diagram, most of the simulated clusters and observed 
open and embedded clusters distribute around a point at which 
the relaxation time and the dynamical time are similar to the
cluster age. This suggests that embedded and open clusters are 
the same population (i.e., embedded clusters are the ancestors of open 
clusters) and that 
they are in the expansion phase after mass segregation.
Observed leaky clusters and some open clusters are located near 
or slightly above the line on which the dynamical time is equal 
to the age, but far away from the line on which the relaxation 
time is equal to the age. They therefore seems to be in an 
evolution phase dynamically different from typical open and 
embedded clusters.

Most of the clusters formed in our simulations evolve to clusters 
similar to open clusters. Some clusters, however, experienced 
hierarchical mergers 
and evolve to dense massive clusters which have characteristics 
similar to those of YMCs in the MW. 
We find that a local star formation efficiency more than $\sim$50\,\%
is necessary for the formation of young massive clusters.
This result is consistent with that of 
\citet{2013A&A...559A..38P}, in which YMCs are suggested to form 
with a SFE of 60--70\%.
If the local SFE is always less than 50\,\%, only open
clusters form instead. This result suggests that the center of 
dense star-forming clumps which evolves to YMCs is almost 
gas-free as is also suggested
by recent simulations \citep{2012MNRAS.419..841K,
2012MNRAS.425..450M}. A local SFE
depending on the local density (i.e., free-fall time) 
\citep{2012ApJ...745...69K,2013MNRAS.436.3167F} is preferable 
for the formation of both open and young massive clusters in 
the same process.

Massive dense clusters formed in our simulation evolved via 
hierarchical mergers within the first few Myr.
Since smaller clumps have a shorter relaxation time,
the merger remnant of smaller clumps shows dynamically 
mature features such as mass segregation,
even if we initially assign the stellar mass irrespective of their 
positions.  
We confirmed that the surface density profile of our simulated
cluster is actually consistent with that of observed YMCs such as 
NGC 3603 and Trumpler 14. We also find that the 
slope of the stellar mass function is shallower in the cluster center 
due to the mass segregation. This feature is also observed 
in young massive clusters in the MW 
\citep{2013ApJ...764...73P,2013AJ....145...46L,2013A&A...556A..26H}.

\bigskip
The author thanks Tsuyoshi Inoue and the anonymous referee for useful 
comments on the manuscript.
This work was supported by JSPS KAKENHI Grant Number 26800108
and NAOJ Fellowship.
Numerical computations were partially carried out on Cray XC30
at the Center for Computational Astrophysics (CfCA) of the National 
Astronomical Observatory of Japan and Little Green Machine 
at Leiden Observatory. 

\bibliographystyle{apj}
\bibliography{reference}

\end{document}